\documentclass[namedreferences]{SolarPhysics}
\usepackage[optionalrh]{spr-sola-addons} % For Solar Physics
\usepackage{epsfig}                     % For eps figures, old commands
\usepackage{natbib}                     % For citations: redefine \cite commands
\newdisplay{guess}{Conjecture}
\newcommand{\ha}{H$\alpha$}

\begin{document}
\begin{article}
\begin{opening}

\title{ARTEMIS IV Radio Observations of the July 14, 2000 Large Solar Event.}
\runningtitle{2000 JULY 14 ARTEMIS IV OBSERVATIONS}
\runningauthor{C. CAROUBALOS  ET AL.}
{
\author{C. \surname{Caroubalos$^1$}}
\author{C. E. \surname{Alissandrakis$^3$}}
\author{A. \surname{Hillaris$^2$}}
\author{A. \surname{Nindos$^3$}}
\author{P. \surname{Tsitsipis}}
\author{X. \surname{Moussas$^2$}}
\author{J-L. \surname{Bougeret$^5$}}
\author{K. \surname{Bouratzis$^2$}}
\author{G. \surname{Dumas$^5$}}
\author{G. \surname{Kanellakis$^6$}}
\author{A. \surname{Kontogeorgos$^4$}}
\author{D. \surname{Maroulis$^1$}}
\author{N. \surname{Patavalis$^1$}}
\author{C. \surname{Perche$^5$}}
\author{J. \surname{Polygiannakis$^2$}}
\author{P. \surname{Preka-Papadema$^2$}}
}
{
\institute{}
\institute{$^1$Department  of Informatics, University of Athens, GR-15783,Athens, Greece}
\institute{$^2$Section of Astrophysics, Astronomy and Mechanics, Department ofPhysics, University of Athens, GR-15784 Athens, Greece}
\institute{$^3$Section of Astro-Geophysics, University of Ioannina, GR-45110 Ioannina, Greece}
\institute{$^4$Department of Electronics, Technological Education Institute of Lamia, Lamia, Greece}
\institute{$^5$Observatoire de Paris, Department de Recherche Spatial, CNRS UA 264, F-92195 Meudon Cedex, France}
\institute{$^6$Thermopylae, O.T.E Inmarsat, GR-35009, Skarpheia, Molos Greece}
}
%\date{May 1, 2001}

\begin{abstract}
In this report we present a complex metric burst, associated with the 14
July 2000 major solar event, recorded by the ARTEMIS-IV radio spectrograph at
Thermopylae. Additional space borne and earthbound observational data are
used, in order to identify and analyse the diverse, yet associated, processes
during this event. The emission at metric wavelengths consisted of broad band
continua, including a moving and a stationary type IV, impulsive bursts and
pulsating structures. The principal release of energetic electrons in the
corona was 15-20 min after the start of the flare, in a period when the
flare emission spread rapidly eastwards and a hard x-ray peak occurred.
Backward extrapolation of the CME also puts its origin in the same time
interval, however the uncertainty of the extrapolation does no allow us to
associate the CME with any particular radio or x-ray signature.
Finally, we present high time and spectral resolution observations of
pulsations and fiber bursts, together with a preliminary statistical analysis.
\end{abstract}

\end{opening}

\section{Introduction}
Radio emission from solar flares can be detected from millimeter to
kilometer wavelengths.  Flare microwave emission  can provide
important diagnostics of acceleration processes in the  corona because
the radio emission is produced by energetic electrons accelerated
during the flare. The microwave flare component is broad--band in
frequency with no appreciable drift and relatively slowly varying
(i.e. with time scales from about tens of seconds up to more than one
hour).

At longer wavelengths, flare radio emission shows an extraordinary
variety of structure in frequency, time and space, some of which is
beyond or at the limit of our observing facilities.  Type III bursts
(cf. McLean, 1985 and references therein) appear on the spectrum as
intense bands of emission drifting rapidly from high to low
frequencies.  They are produced as electrons, accelerated at low
altitudes, gain access to open field lines and propagate well out into
the corona.  Type U  bursts on the other hand trace the propagation of
energetic  electrons along closed magnetic field lines (coronal loops).

Type II bursts represent the passage of an MHD  shock wave through the
tenuous plasma of the solar corona; their radio emission is  due
either to energetic electrons accelerated at the shock front or plasma
turbulence  excited by the shock; they originate either by a flare
blast wave or by a CME forward shock (Maia et al. 2000, Aurass 1997).

The continua observed during periods of activity, on the other hand,
represent the radiation of energetic electrons trapped within
magnetic clouds, CMEs and plasmoids and they appear under the name
type IV bursts. The moving type IV bursts  (Boischot 1957, also dubbed
IV mA after Takakura 1969; see also Slottje, 1981) are emitted from
sources of meter wave continuum which are believed to move  outwards
at velocities of the order of 100-1000 km/sec; their spectrum is often
featureless and sometimes last more than 10 minutes. Some of them
appear following type II bursts and  are possibly caused by energetic
electrons produced in the wake of the type II shock. Others originate
from energetic electron populations trapped in  expanding magnetic
arches or plasmoids, i.e. blobs of dense plasma containing their
own magnetic field (cf. Wild and Smerd 1972; Stewart, 1985 and
references therein).  The type IV mA bursts which are believed to
originate within CMEs may be used for the CME detection by radio
methods (Aurass 1997).  The stationary type IV bursts (IV mB after
Takakura 1969; see also Slottje, 1981)  emanate from stationary sites
usually located above active regions (Robinson, 1985). Type IV events
present fine  structure, such as pulsating structures, fiber bursts
and zebra patterns, while much less  abundant are spike bursts,
braided zebra patterns and the extremely rare tadpole  bursts
(Slottje, 1981).

\begin{figure}
\centerline{\epsfig{file=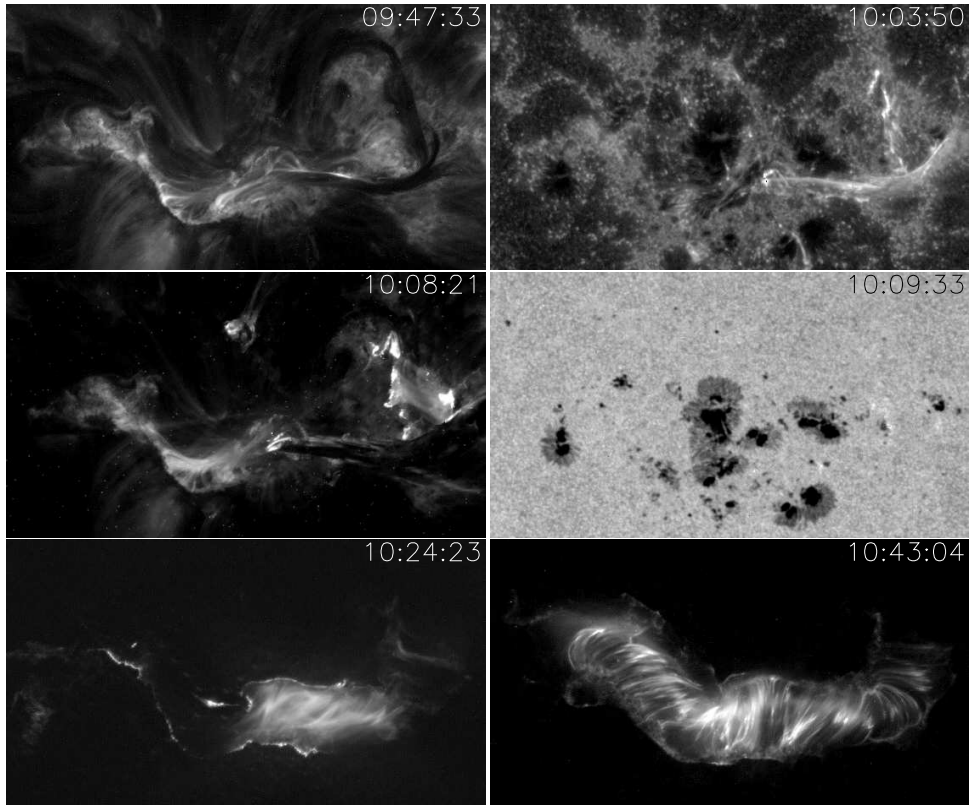,width=\textwidth}}
\caption{TRACE images before and during the flare. The image at 10:03:50 UT
is at 1600 \AA, the one at 10:09:33 is in white light, the others are at 195 \AA.}
\end{figure}

In this article we present a preliminary analysis of a complex event that took
place on July 14, 2000. It produced a high flux of high energy particles that
reached the earth (24000 proton flux units, reported by NOAA Space Environment
Centre), while the associated of the solar wind velocity measured by various
spacecraft near Earth was about 1000 km/s. The flare was accompanied by a
filament eruption and a coronal mass ejection. A multitude of space borne and
ground based instruments observed the event in the entire electromagnetic
spectrum from kilometric radio waves to hard X-rays. Our analysis is based on
observational data obtained with the ARTEMIS-IV multichannel radio
spectrograph, studied in conjunction with images from the Nan\c cay
Radioheliograph as well as data from soft and hard X-ray instruments.

Section 2 presents the observations. The analysis of the data, covering both
the global aspects of the burst evolution and the fine structure is presented
in section 3. Discussion and conclusions are given in section 4.

\begin{figure}
\epsfig{file=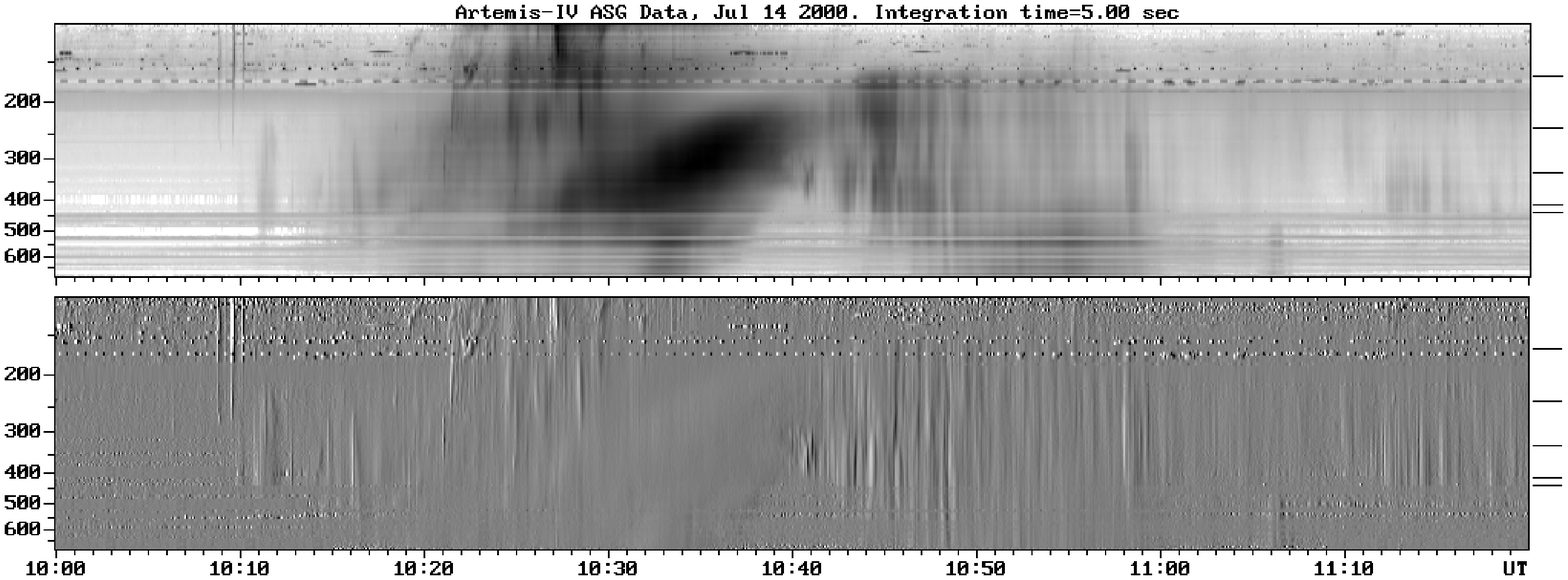,width=12.5cm}

\hspace{-0.68cm}
\epsfig{file=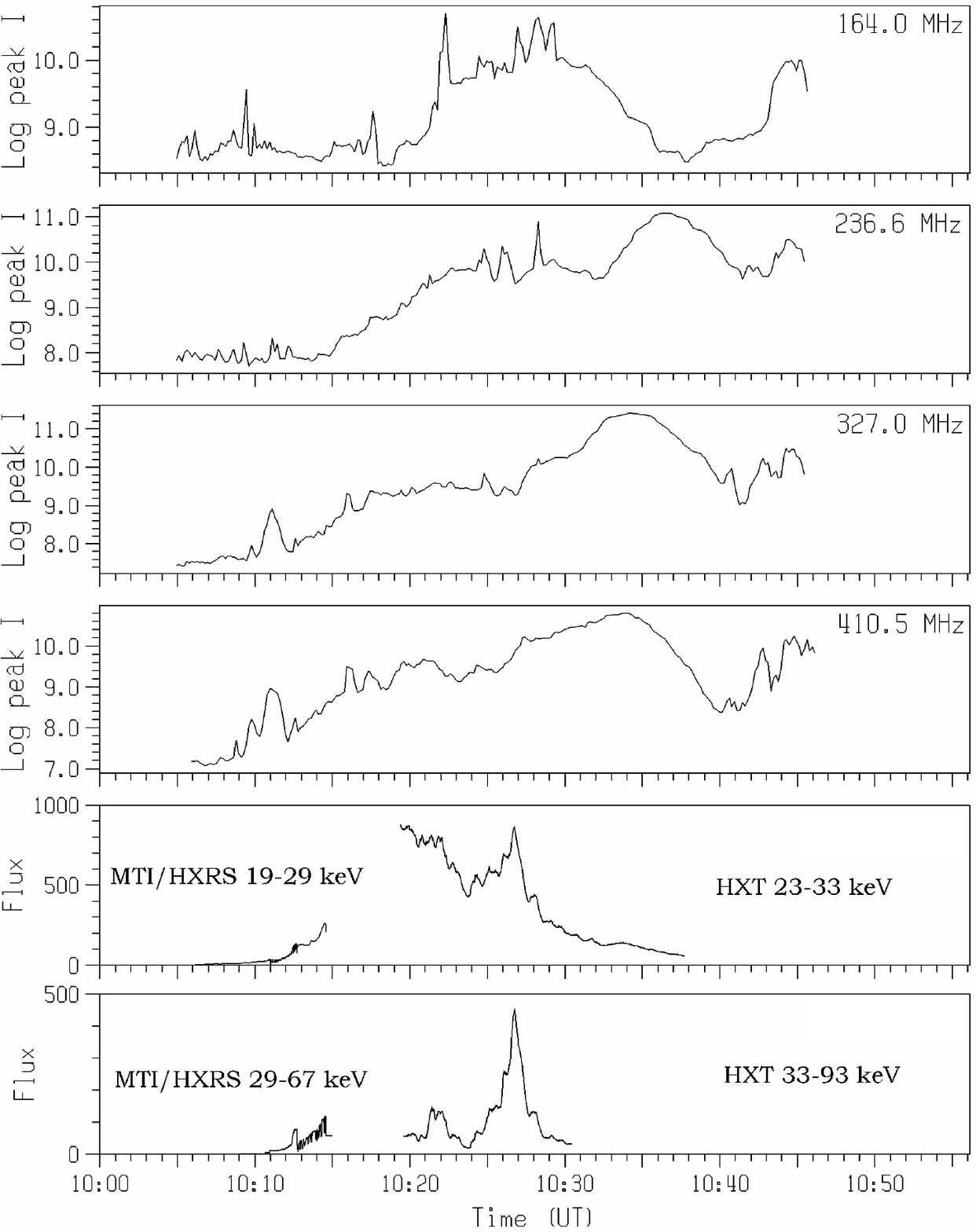,width=9.423cm}
\caption{The dynamic spectrum of the event observed with the ARTEMIS-IV ASG
receiver (upper two panels); the lower panel shows the time derivative of the
flux, which helps to identify rapidly varying structures. Horizontal lines
at the right of the spectrum mark the frequencies of the Nan\c cay
Radioheliograph channels. The four plots below the spectrum give the logarithm
of the peak intensity (in K), measured by the NRH at four frequencies. The
lower two panels show the hard X-ray flux observed with the Yohkoh SXT and
the MTI HXRS instruments.}
\end{figure}

\section{Observations}
Our observations were obtained with the ARTEMIS-IV solar radiospectrograph,
operating at Thermopylae, Greece (Caroubalos et al 2000). The instrument
covers  the frequency range of 110 to 687 MHz, using two receivers operating
in parallel. One, the Global Spectral Analyser (ASG), is a sweep frequency
receiver and the other, the Acousto-Optic Spectrograph (SAO), is a
multichannel acousto-optical receiver. The sweep frequency analyser covers
the full frequency band with a time resolution of 10 samples/s, while the
high sensitivity multi-channel acousto-optical analyser covers the 265-450
MHz range, with high frequency and time resolution (100 samples/s).
In the present study we exploit both data sets: The broad band, medium time
resolution set of the ASG in order to study the association of spectral
features of the event with 2D images from the Nan\c cay Radioheliograph (NRH)
as well as with data from other instruments and the narrow band, high time
resolution SAO mostly for the analysis of the fine temporal and spectral
structures.

The Nan\c cay Radioheliograph provides two-dimensional images of the sun
at five frequencies between 450 and 150 MHz with sub-second time resolution
(Bonmartin et al 1983, Kerdraon and Delouis 1997). For the present study we
used images integrated over 10 sec, between 10:05 and 10:45 UT (provided by
the Radioheliograph group), at 164, 236, 327, 410.5 and 432 MHz; All five
frequencies are within the spectral range of the ASG, while the last three
are also within the range of the SAO. Radio emission at these frequencies
comes from the low and middle corona (height 0.1-0.5 solar radii).

In addition to the Artemis and NRH data, we made use of soft X-ray
observations from TRACE (at 195 \AA\ and 1600 \AA), GOES and the Yohkoh SXT
instruments. The
hard X-ray data were from the Yohkoh HXT and the HXRS aboard MTI (Farnik et al
1998). The former is a Fourier synthesis type imager with 0.5 s time
resolution in four energy bands between 14 and 93 keV. The latter is an X ray
spectrometer with 200 ms time resolution in flare mode in the range 12.6 to
250 keV. Finally we used white light coronograph data from LASCO C2 and C3
coronographs.

\section{Results}
\subsection{Overview}
Descriptions of the event have been given by Zhang et al (2001), Klein et al
(2001) and Karlicky et al (2001). We summarize here and in Table I the
main stages of its evolution, in conjunction with the radio spectral data.

\begin{figure}
\centerline{\epsfig{file=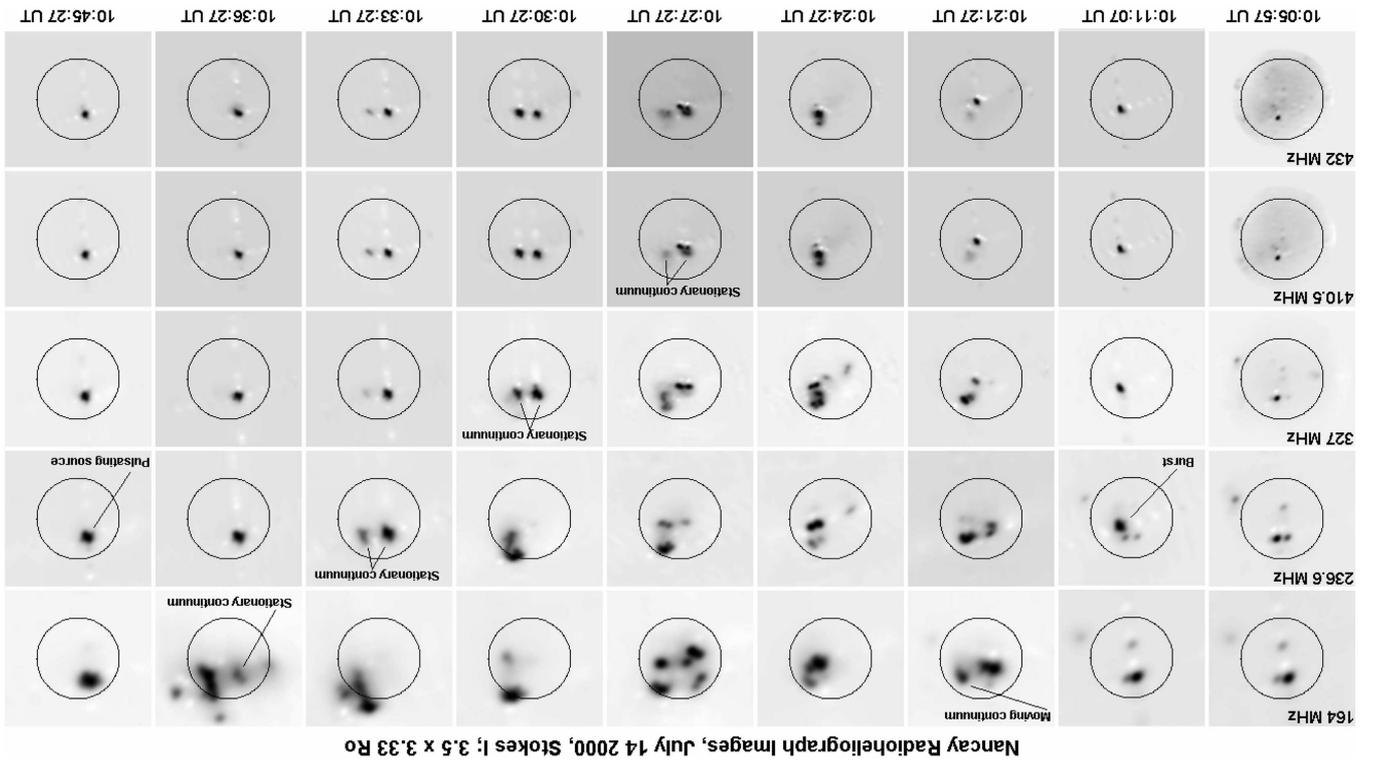,width=10cm,angle=90}}
\caption{Images at five radio frequencies obtained with the Nan\c cay
Radioheliograph, illustrating the evolution of the event. Each image row
corresponds to a particular frequency and each column to a particular time.}
\end{figure}

The H$\alpha$ flare started at 10:03 UT, with a maximum at 10:24, in active
region 9077 near the disk center (N17E01). The GOES flux started to rise at
10:03:06 UT in the 3-25 keV channel and 50 s later in the 1.5-10 keV channel;
the maximum was reached at 10:24 UT and the event was classified as X5.7.
There is a data gap in the TRACE 195 \AA\ images in the critical
time interval between 09:58 and 10:08 UT, therefore we used 1600 \AA\
continuum images in order to detect the flare onset. The first flare kernels
 were visible in
the image at 10:03:50 UT (Figure 1), which is consistent with the GOES data.
Before that time there was some activity along the neutral line filament.
At 10:08:21 UT the flare ribbons
were already well visible in the NW part of the active region, with bright
emission between them, apparently from diffuse loops. In the same TRACE image the disrupted filament is clearly seen;
the filament the filament is seen to rise rapidly and can be followed till
about 10:30 UT, silueted dark in front of the bright flaring region.

The flare was confined to the west part of the active region until about
10:24 UT, at which time it spread abruptly to the east part (Figure 1).
At about the same
time distinct fine loop structures joining the flare ribbons became prominent
in  the west part of the flare, while by 10:40 UT the entire flaring region,
both in the west and the eastern part of the region was dominated by a
spectacular loop system

In Figure 2 (lower panels) we present HXT data in the energy range
of 23-33 keV and 33-93 keV, combined with HXRS data in the 19-29, 29-67.2 keV
range respectively. The combined Yohkoh HXT and HXRS data leave a gap between
10:15 and 10:19:36 UT, within which apparently a significant part of the hard
x-ray emission occurred. After the gap there is a peak at 10:26:45 UT, which
is harder than the rest of the observed part of the burst emission and
coincides in time with the spread of the flare in the west part of the active
region. According to HXT imaging, it consists of two compact components. We
could not reliably determine their exact position because they were located
outside the SXT field of view, which included only the west part of the
flaring region; we can only say that the components  were east of the
SXT field of view, which brings them close to region where the flare expanded.

\begin{figure}
\centerline{\epsfig{file=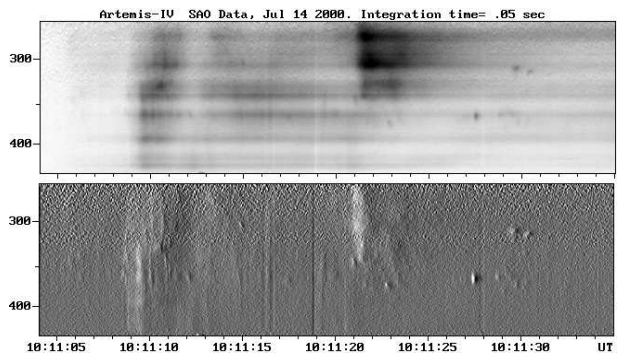,width=8cm}}
\caption{High resolution dynamic spectrum of the burst shown in the second
column of Figure 3, observed with the SAO, in flux (top) and differential
(bottom) displays}
\end{figure}

The dynamic spectrum of the event, observed with the ASG, is given in the upper
panel of Figure 2. The temporal resolution has been reduced to 5 sec, in
order to provide an overall view of the event. In the same figure we give
logarithmic plots of the peak intensity observed with the Radioheliograph at
four frequencies. Figure 3 presents selected
images from the Nan\c cay Radioheliograph that, in conjunction with the
dynamic spectrum, illustrate the evolution of the event.

\begin{table}{}
\caption{{\em Major phases of the July 14 event}}
\begin{tabular}{|l|l|l|}
\hline
Time (UT)& Optical \& X-rays                 & Metric Radio\\
\hline
10:03 &Start of flare in H$\alpha$ \& soft X-rays&Pre-existing noise storms\\
10:06 &                                      &Weak continuum onset at 432 \& 411 MHz\\
10:07:36 &                                   &Weak continuum onset at 327 MHz\\
10:08-10:10 &                                & Noise storm associated type III's \\
10:08:21 & Flare \& disrupted filament       &\\
         & clearly visible at 195 \AA\       &\\
10:09:27 &                                   &Weak continuum onset at 327 MHz\\
10:11:10 &                                   &First flare associated drift bursts \\
10:12    &                                   &Continuum seen in dynamic spectrum\\
10:15:37 &                                   &Weak continuum onset at 164 MHz\\
$\sim$10:18&                                 &Start of motion of continuum source\\
10:20-10:30&                                 &Isolated type III and III G's\\
10:24    & Flare maximum (GOES)              &\\
10:24:20 & Flare spreads East                 &\\
10:26:45 & Hard X-ray peak $>$33keV          & \\
10:27    &                                   & Onset of stationary continuum\\
10:30    & Filament last seen at 195 \AA\    &              \\
10:34:00 &                                   & Continuum peak at 411 MHz\\
10:34:30 &                                   & Continuum peak at 327 MHz\\
10:36:30 &                                   & Continuum peak at 237 MHz\\
10:38    &                                   & Onset of fiber bursts \& pulsations\\
10:41    &                                   & End of stationary continuum\\
10:54    & Halo CME reaches 5.4 $R\odot$ in C2 & \\
         & ($v$=1700 km/s)                     & \\
11:18    & Halo CME reaches 9.4 $R\odot$ in C3 & \\
         & ($v$=3310 km/s)                     & \\
11:42    & Halo CME reaches 12.5 $R\odot$ in C3& \\
         & ($v$=1000 km/s)                     & \\
\hline
\end{tabular}
\end{table}

The dynamic spectrum shows a complex pattern of emission, which inclu\-des
impulsive bursts, continua and pulsations. The Radioheliograph group (Klein
et al 2001) reported noise storms preceding the flare, located near the
central meridian in the Northern Hemisphere (Figure 3, left column). It is
interesting to note that there was very little
other activity until several minutes after the start of the flare. The group
of type III's around 10:09 UT (Figure 2) is associated with the noise storm
rather than with the flaring region. Another group of what appear to be fast
drift bursts in the ASG spectrum occurred around 10:11 UT at frequencies
above 210 MHz. The corresponding
NRH images are shown in Figure 3, second column; they are visible at all NRH
frequencies except 164 MHz and are located near the flare. For comparison, we
give in Figure 4 the dynamic spectrum of this event with 50 ms time resolution,
observed with the SAO. The high resolution observations reveal considerable
fine structure and practically no frequency drift of the individual components.

\begin{figure}
\centerline{\epsfig{file=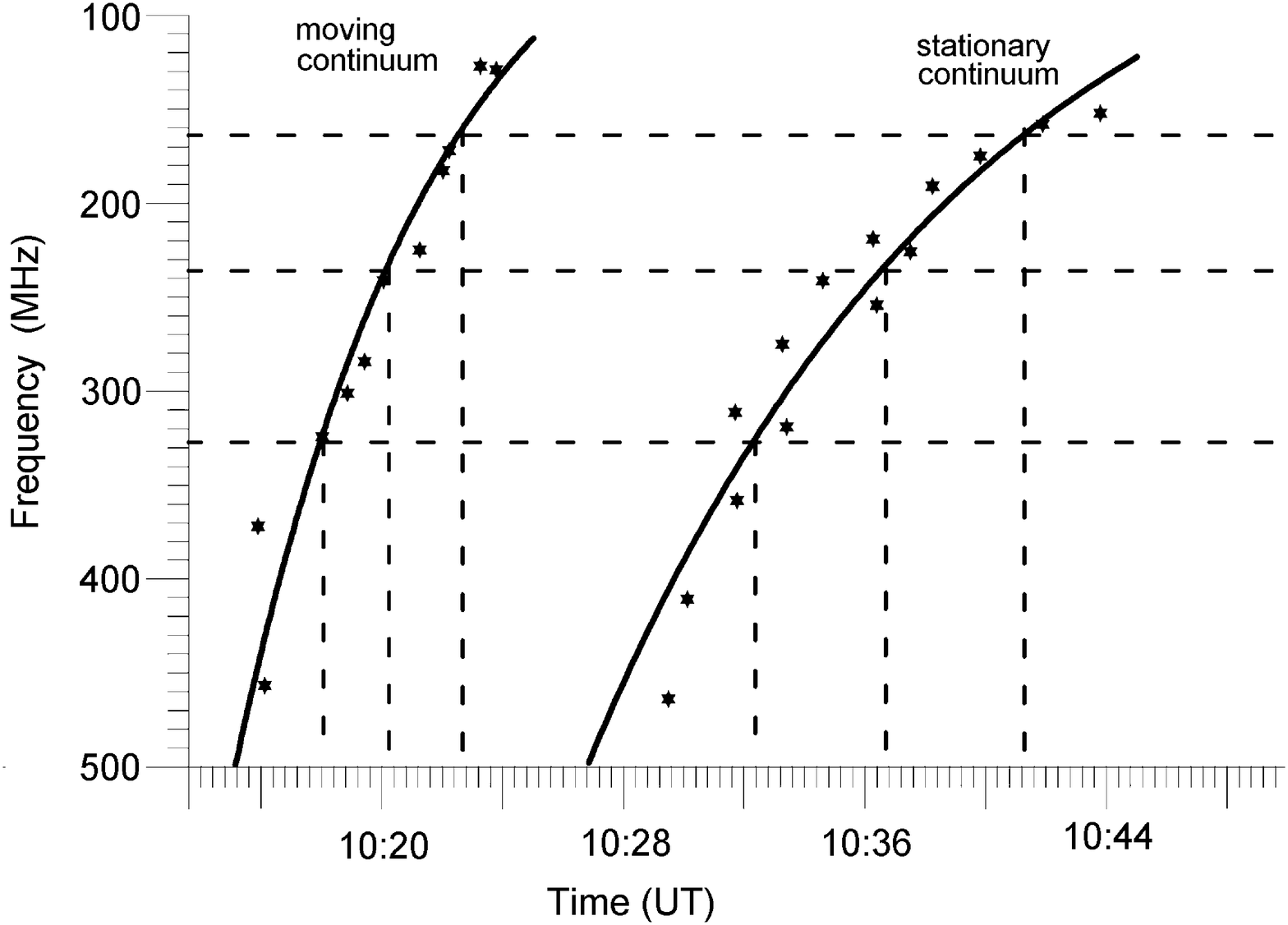,width=8cm}}
\caption{Position of maxima of the moving and stationary continua
as a function of frequency (linear scale) and time. Both are fitted with
exponential curves (solid lines). Horizontal dashed lines mark the 164, 236.5
and 327 MHz levels of NRH, vertical lines mark the time at which each
continuum crosses the corresponding layer.}
\end{figure}

Impulsive bursts, including a type III storm as well as short scale time
variations, became more frequent in the time interval 10:20 to 10:30 UT,
gradually drifting to lower frequencies (Figure 2). It is interesting to note
that they occurred around the time of the hard x-ray peak (10:26:45 UT)
and the spread of the flare in the east part of the active region. Therefore
it appears that, before this time, very few energetic particles could find
their way to the corona, although such particles were released in the lower
atmosphere as evidenced by the hard X-ray emission.

The fast drift bursts coexist with a slowly drifting continuum, which in
the dynamic spectrum became prominent around 10:11 UT at high frequencies and
was detectable until about 10:39 UT at the low frequency end of the ASG band.
Even before 10:11, continuum emission appeared at 10:06 UT in the 432 and
410.5 MHz NRH images, as a localized source at the west side of the flaring
region. Tis emission became later visible at 327 MHz ($\sim$10:07:37 UT), at
236.5 MHz ($\sim$10:09:27 UT) and finally at 164 MHz ($\sim$10:15:37 UT);
clear evidence for its existence and progressive appearance from high to low
frequencies is provided by the logarithmic NRH intensity time curves of
Figure 2.

\begin{figure}
\centerline{\epsfig{file=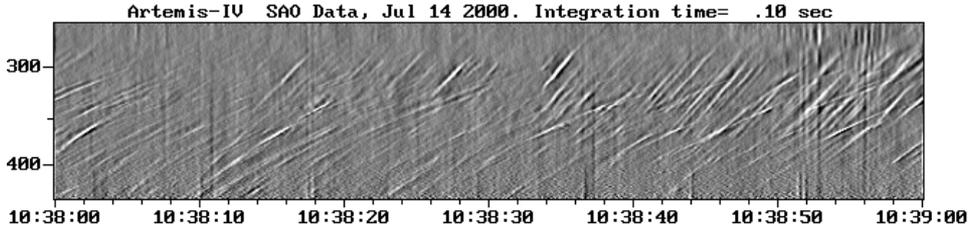,width=\textwidth}}
\caption{Time derivative of the dynamic spectrum observed with the SAO over
a one minute interval in the 265-441 MHz range}
\end{figure}

The position of the continuum peak on the dynamic spectrum is shown in Figure
5 (left curve). Although its drift rate is within the range of type II bursts,
Klein et al (2001) pointed out that it had too large a banwidth and lacked
fine spectral structure. This slow drift continuum was practically stationary
until about 10:18 UT, when part of the emission started moving north, in the
form of a moving type IV (Klein et al 2001; cf Figure 3, columns 3-6)

The most prominent feature of the dynamic spectrum is a broad, structureless
continuum, apparently a type IV, which started around 10:27 UT and extended
above 1.5 GHz (see
Figure 10 of Karlicky et al 2001). Although it shows a clear frequency drift
(Figure 5, right curve) it was associated with two stationary sources
(Figure 3, columns 5-8), one located west and the other east of the active
region. The western source faded around 10:35, while the eastern source
persisted until 10:41 UT and was followed by pulsations, which originated at
the same location (Figure 3, last column). The pulsations show a very complex
temporal/spectral structure in the ASG spectrum, while they are much better
visible in the ASG spectrum (see next section). They appear at frequencies
above 150 MHz and persisted until about 11:30 UT.

The flare was accompanied by an Earth-directed halo coronal mass ejection
(CME). The analysis of the SOHO coronograph (C2, C3) images shows that the
velocity of the CME, estimated on the plane of the sky, was about 1800 km/s,
with lower and upper limits of 1660 and 1900 km/s respectively. The large
expansion velocity of the CME between 5.4 and 9.4 $R_{\odot}$ (Table I) might
be attributed to errors of estimation of the extent of the CME, but it is also
possible that an acceleration of the CME by the vast number of particles
injected by the explosive phenomena during this time period took place.

\begin{figure}
\centerline{\epsfig{file=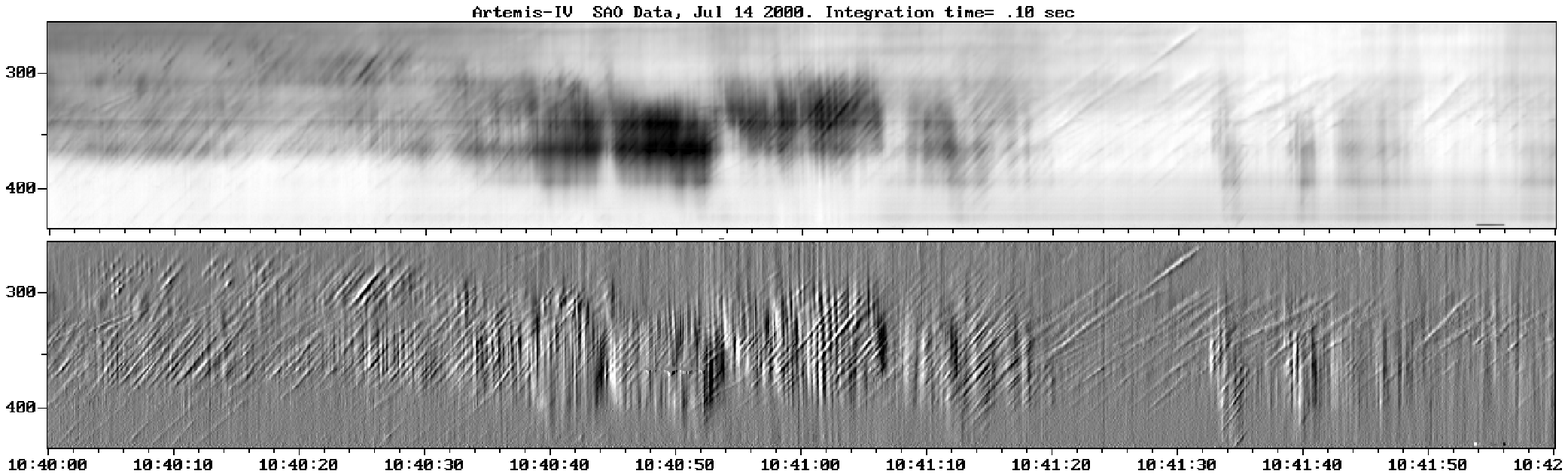,width=12.5cm}}
\centerline{\epsfig{file=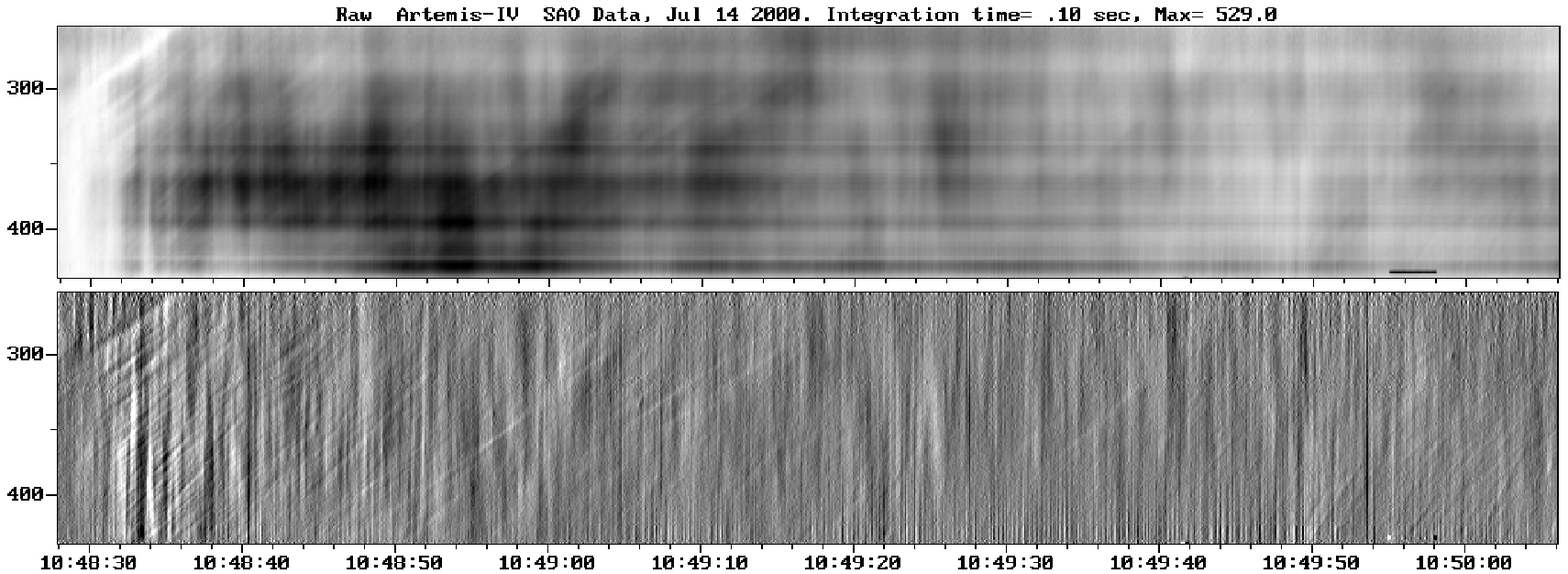,width=10.24cm}}
\centerline{\epsfig{file=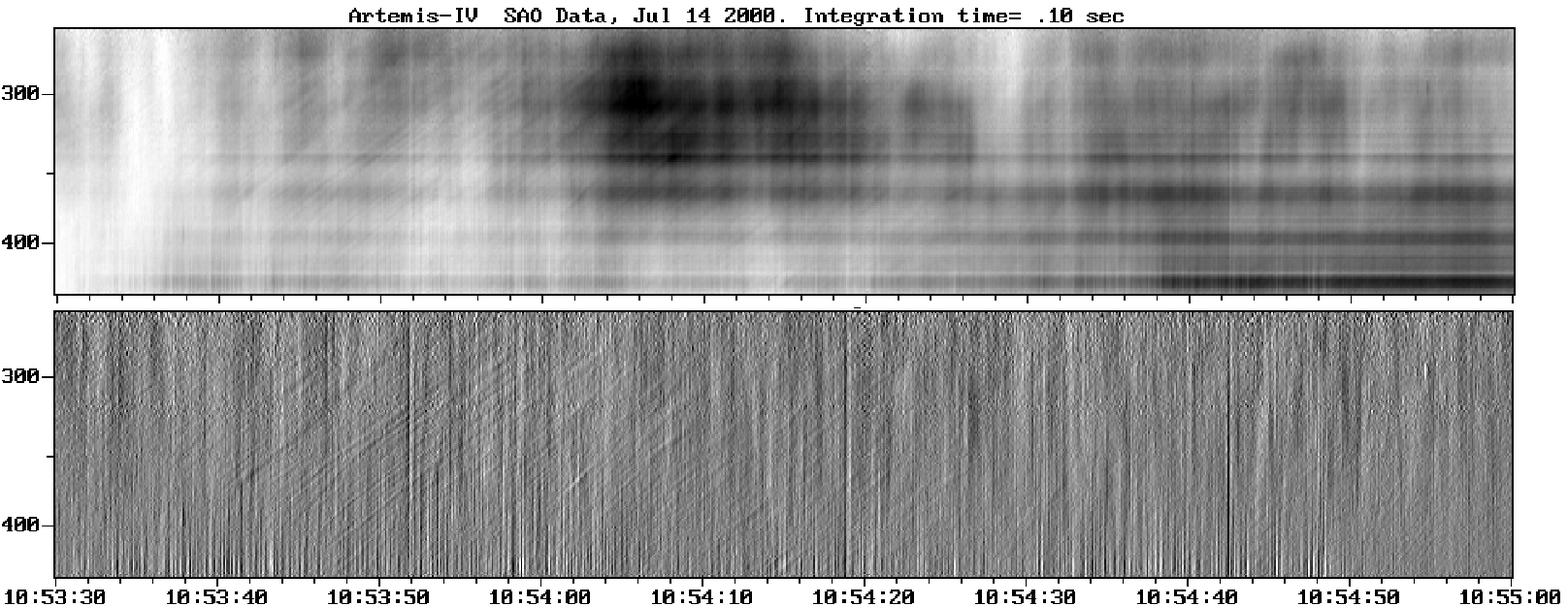,width= 9.66cm}}
\caption{Dynamic spectra of fiber bursts and pulsations (flux and its time
derivative) during the late stage of evolution of the event.}
\end{figure}

Backwards extrapolation of the CME position, on the basis of the measurements
given in table I, gives that the CME exploded near 10:23:20 UT ($\pm$7.5 m).
Whatever the value of this raw estimate, it gives evidence that the origin of
the CME is associated with the enhanced activity in the interval 10:20-10:30
UT described above, although it is impossible from this information alone  to
establish its association with and of the metric radio components.

\subsection{Fine Structure}
Due to its high sensitivity, spectral and temporal resolution, the SAO is
ideal for studies of fine temporal structure, such as the one following the
stationary continuum. In this section we present examples of such structures,
together with a preliminary analysis.

A one minute long dynamic spectrum in the interval 10:38 to 10:39 UT is shown
in Figure 6; what is actually shown is the time derivative of the flux, which
eliminates any underlying continuum emission and brings up fast varying
structures. The most prominent feature in the spectrum is the fine, drifting
bursts (fiber bursts) of $\sim$300\,ms duration. They appear at frequencies
above 280 MHz and at least two families with different drift rates coexist.

Three more examples are shown in figure 7, at different periods during the
late stages of evolution of the event. Both the flux and its time derivative
are shown; in the flux spectrum, the average value has been subtracted in
order to show better the fine structure.

The top spectrum shows a mixture of fiber bursts and pulsations, in the
interval 10:40-10:42 UT. The fiber bursts have similar characteristics to
those of Figure 6; the pulsations, with periods of $\sim$200\, ms,
were limited into the spectral range 290-410 MHz, and appeared in groups.
Later in the event (third spectrum in Figure 7), the pulsations became more
numerous and extended beyond the 265-445 MHz SAO band; they became the
predominant feature still later (bottom spectrum of Figure 7).

We are in the process of developing numerical methods for studying the
statistical properties of the numerous fine features seen in the spectra. We
have already developed a feature tracking algorithm, on the basis of which we
can compute the distribution of the drift rates of fiber bursts. An example is
shown in Figure 8, where we give the histogram of drift rates for the same
interval as in Figure 6. As expected, most features show negative frequency
drifts and the two fiber burst families of Figure 6 are discernible, associated
to histogram peaks at $-6$ and $-12$ MHz/s; the histogram also shows a few
reverse (positive) drift structures.

\begin{figure}
\centerline{\epsfig{file=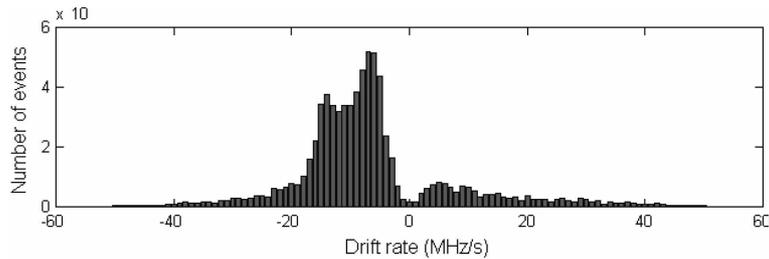,width=10cm}}
\caption{Distribution of drift rates during the interval of 10:38-10:39 UT
(cf Figure 6)}
\end{figure}

\section{Summary and Conclusions}

The event of July 14, 2000 is typical of a large flare, manifesting itself
in all possible ways. The flare produced complex metric radio emission,
with several components: impulsive, continua and pulsations. Imaging data
from the Nan\c cay Radioheliograph, used in conjunction with spectral data
from Artemis-IV, facilitated their identification.

Although the flare started at 10:03 UT in \ha\ and soft x-rays, it was the
time interval 10:18 to 10:30 that was the most rich in phenomena, including
a moving type IV followed by a stationary type IV, a hard x-ray peak and
a rapid expansion of the flare eastwards. It also appears that during this
period most of the energetic electrons were released in the corona, as
evidenced by the multitude of fast drift bursts. Backwards extrapolation of
the CME position measured on LASCO C2 and C3 images also points to this time
interval for the origin of the CME, although its association with any
particular feature of the radio emission is not clear.

The metric emission was also rich in pulsating structures and fiber bursts,
which followed the stationary continuum and lasted for more than one hour. We
presented a preliminary statistical analysis of fiber bursts, that reveals two
co-existing families with different drift rates. We are in the process of
improving our analysis and we expect to present more thorough results in the
near future.

\begin{acknowledgements}
The authors are indebted to the Meudon Radioheliograph Group and K.-L. Klein
for providing the Nan\c cay Radioheliograph images; they are also grateful to
Frantisek Farnik who provided the MTI/HXRS data. This work was supported in
part by a grant from the University of Athens.
\end{acknowledgements}

\end{article}
\end{document}